\documentclass[%
 reprint,
 superscriptaddress,
 amsmath,amssymb,
 aps,
]{revtex4-2}

\usepackage{graphicx}
\usepackage{dcolumn}
\usepackage{bm}
\usepackage{hyperref}
\usepackage[mathlines]{lineno}
\usepackage{color}
\usepackage{siunitx}
\usepackage{subfigure}
\usepackage{xspace}
\usepackage{orcidlink}
\usepackage{soul}

\DeclareSIUnit[quantity-product = ]\percent{\char`\%}

\usepackage{soul}
\usepackage{xcolor}

\begin{document}

\title{\boldmath Probing KSVZ Axion Dark Matter near 5.9\,GHz Using a 8-Cell Cavity Haloscope}

\author{Saebyeok Ahn}
\thanks{These authors contributed equally to this work.}%
\affiliation{Dark Matter Axion Group, Institute for Basic Science, Daejeon 34126 Republic of Korea}
\affiliation{Center for Axion and Precision Physics Research, Institute for Basic Science (IBS), Daejeon 34051, Republic of Korea}

\author{\c Ca\u glar Kutlu}
\thanks{This author also contributed equally to this work.}
\altaffiliation[Present address: ]{Zurich Instruments, Technoparkstrasse 1, 8005 Z\"urich, Switzerland}
\affiliation{Center for Axion and Precision Physics Research, Institute for Basic Science (IBS), Daejeon 34051, Republic of Korea}
\affiliation{Department of Physics, Korea Advanced Institute of Science and Technology (KAIST), Daejeon 34141, Republic of Korea}

\author{Soohyung Lee\,\orcidlink{0000-0001-5959-9407}}
\thanks{These authors contributed equally to this work.}
\affiliation{Center for Axion and Precision Physics Research, Institute for Basic Science (IBS), Daejeon 34051, Republic of Korea}
\affiliation{Center for Accelerator Research, Korea University, Sejong 30019, Republic of Korea}

\author{SungWoo Youn}
\thanks{Corresponding author}
\email{swyoun@ibs.re.kr}
\affiliation{Dark Matter Axion Group, Institute for Basic Science, Daejeon 34126 Republic of Korea}
\affiliation{Center for Axion and Precision Physics Research, Institute for Basic Science (IBS), Daejeon 34051, Republic of Korea}

\author{Sungjae Bae}
\affiliation{Dark Matter Axion Group, Institute for Basic Science, Daejeon 34126 Republic of Korea}
\affiliation{Center for Axion and Precision Physics Research, Institute for Basic Science (IBS), Daejeon 34051, Republic of Korea}
\affiliation{Department of Physics, Korea Advanced Institute of Science and Technology (KAIST), Daejeon 34141, Republic of Korea}

\author{Junu Jeong}
\affiliation{Center for Axion and Precision Physics Research, Institute for Basic Science (IBS), Daejeon 34051, Republic of Korea}
\affiliation{Oskar Klein Centre, Department of Physics, Stockholm University, AlbaNova, SE-10691 Stockholm, Sweden}

\author{Arjan F. van Loo}
\affiliation{RIKEN Center for Quantum Computing (RQC), Wako, Saitama 351-0198, Japan}
\affiliation{Department of Applied Physics, Graduate School of Engineering, The University of Tokyo, Bunkyo-ku, Tokyo 113-8656, Japan}

\author{Yasunobu Nakamura}
\affiliation{RIKEN Center for Quantum Computing (RQC), Wako, Saitama 351-0198, Japan}
\affiliation{Department of Applied Physics, Graduate School of Engineering, The University of Tokyo, Bunkyo-ku, Tokyo 113-8656, Japan}

\author{Seongjeong Oh}
\affiliation{Dark Matter Axion Group, Institute for Basic Science, Daejeon 34126 Republic of Korea}
\affiliation{Center for Axion and Precision Physics Research, Institute for Basic Science (IBS), Daejeon 34051, Republic of Korea}

\author{Sergey V. Uchaikin}
\affiliation{Dark Matter Axion Group, Institute for Basic Science, Daejeon 34126 Republic of Korea}
\affiliation{Center for Axion and Precision Physics Research, Institute for Basic Science (IBS), Daejeon 34051, Republic of Korea}

\author{Jihn E. Kim}
\affiliation{Department of Physics, Seoul National University, Seoul 08826, Republic of Korea}

\author{Yannis K. Semertzidis}
\affiliation{Center for Axion and Precision Physics Research, Institute for Basic Science (IBS), Daejeon 34051, Republic of Korea}
\affiliation{Department of Physics, Korea Advanced Institute of Science and Technology (KAIST), Daejeon 34141, Republic of Korea}

\date{\today}

\begin{abstract}
We report on a search for axion dark matter in the frequency range near 5.9\,GHz, conducted using the haloscope technique. 
The experiment employed an 8-cell microwave resonator designed to extend the accessible frequency range by a multi-fold factor relative to conventional single-cell configurations, while maintaining a large detection volume. 
To enhance sensitivity, a flux-driven Josephson parametric amplifier (JPA) operating near the quantum noise limit was utilized, together with a sideband-summing method that coherently combines mirrored spectral components generated by the JPA.
Data were acquired over the frequency range 5.83--5.94\,GHz. 
With no statistically significant excess observed, we exclude axion-photon couplings $g_{a\gamma\gamma}$ down to $1.2 \times 10^{-14}$\,GeV$^{-1}$ at a 90\% confidence level. 
The achieved sensitivity approaches the KSVZ benchmark prediction, setting the most stringent limits to date in this range.
\end{abstract}

\maketitle

A well-motivated solution to the strong $CP$ problem in the Standard Model of particle physics~\cite{PRL37_8_1976,PRD14_3432_1978,PRD18_2199_1978,PR108_120_1957,PRD15_9_1977,NPA341_269_1980} is the so-called Peccei-Quinn mechanism~\cite{PRL38_1440_1977}. 
It introduces a new global U(1) symmetry, which is spontaneously broken and dynamically relaxes the CP-violating term in the quantum chromodynamics Lagrangian.
The spontaneous symmetry breaking gives rise to a pseudo-Nambu–Goldstone boson known as the axion~\cite{PRL40_223_1978, PRL40_279_1978}.
The axion is also a leading candidate for cold dark matter, which constitutes approximately 85\% of the matter content of the present universe~\cite{AA594_A13_2016}, yet remains unidentified. 

The haloscope is one of the most promising experimental strategies to detect the axion in our galactic halo~\cite{PRL51_1415_1983}.
It relies on the axion’s electromagnetic interaction, enabling axion-to-photon conversion in the presence of a static magnetic field. 
Microwave cavities are often employed to resonantly enhance the photon signal.
The axion-photon coupling is given by $g_{a\gamma\gamma}=\frac{\alpha g_{\gamma}}{\pi {f_{a}}}$, where $\alpha$ is the fine structure constant, $f_a$ is the axion decay constant and $g_{\gamma}$ is a model-dependent dimensionless coefficient. 
In the most popular models, Kim-Shifman-Vainshtein-Zakharov (KSVZ)~\cite{PRL43_103_1979,NPB166_493_1980} and Dine-Fischler-Srednicki-Zhitnitskii
(DFSZ)~\cite{YF31_497_1980,PLB104_199_1981}, $g_{\gamma}$ takes values of $-0.97$ and 0.36, respectively.


The axion-induced photon signal power, maximized when the axion frequency $\nu_a$ (corresponding to mass $m_a=2\pi\nu_a$) matches the resonant frequency $\nu_c$ of the cavity, can be expressed as
\begin{equation}
\begin{split}
	P_{a\gamma\gamma} &= 14.2\,{\rm yW} \left(\frac{g_{\gamma}}{0.97}\right)^2 \left(\frac{\rho_{a}}{0.45\,{\rm GeV}}\right) \left(\frac{\nu_a}{5.9\,{\rm GHz}}\right) \\
	& \times\left(\frac{B_{\rm rms}}{6.97\,{\rm T}}\right)^{2} 
    \left(\frac{V}{3.1\,{\rm L}}\right)\left(\frac{C}{0.6}\right)\left(\frac{Q_{L}}{15000}\right),
\end{split}
\label{eq:axion_conversion_power}
\end{equation}
where $\rho_{a}$ is the local axion dark matter density, $B_{\rm rms}$ is the root-mean-square magnetic field inside the cavity volume $V$, $C$ and $Q_L$ are the form factor and loaded quality factor of the cavity mode considered for the experiment, respectively. 

For cylindrical resonators, the lowest fundamental TM mode, TM$_{010}$, is typically chosen due to its maximal form factor under a uniform solenoidal magnetic field, thereby optimizing sensitivity to axion-induced signals. 
However, the cavity dimensions are generally constrained by the magnet bore, limiting conventional haloscope designs to relatively low-frequency, and thus low-mass axion searches. 
To overcome this limitation, we implement a novel multi-cell cavity architecture that enables operation at higher resonant frequencies while maintaining a large detection volume~\cite{plb777_412_2018, prl125_221302_2020}. 
In this Letter, we report on a haloscope search for KSVZ axion dark matter near 5.9\,GHz (corresponding to a mass of 24.4\,$\mu$eV), employing an 8-cell microwave cavity for high-frequency access and a flux-driven Josephson parametric amplifier (JPA) for quantum-limited signal readout.

\begin{figure}
	\centering
        \includegraphics[width=0.9\linewidth]{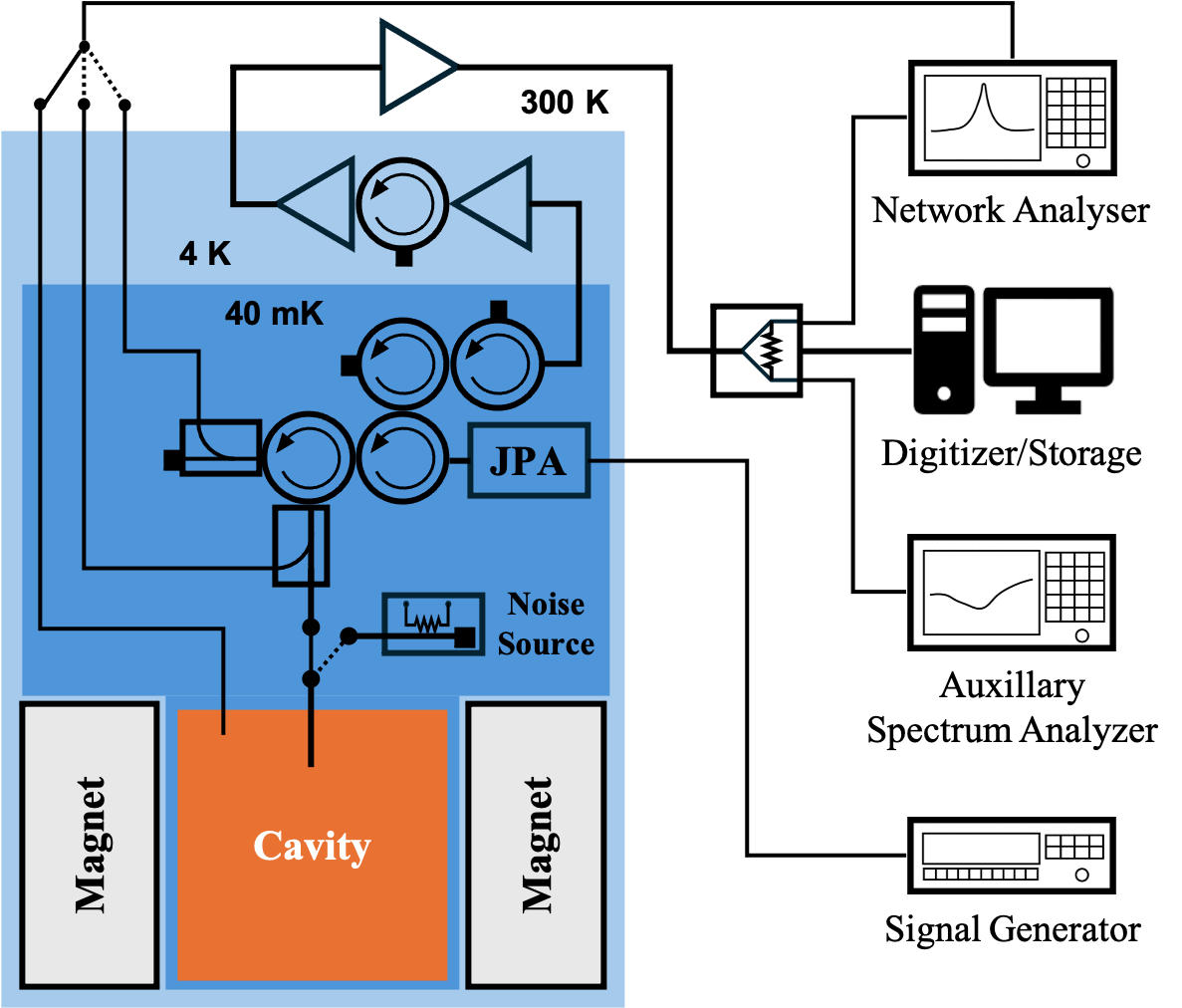}
		\caption{Schematic diagram of the experimental setup.}
        \label{fig:system_overview}
\end{figure}

The experimental setup featured 1) a superconducting magnet with a 165-mm bore, capable of generating a central magnetic field of up to 8\,T, and 2) a dilution refrigerator that provides the required cryogenic environment--4\,K for the magnet and below 40\,mK for the detector components, including the resonant cavity and the JPA.
The resonant cavity, symmetrically positioned within the magnet bore, served as the detection volume for axion conversion.
Signal amplification was initiated by the JPA and followed by two high-electron-mobility transistor (HEMT) amplifiers. 
The amplified signal was then routed through a chain of room-temperature electronics, including an additional HEMT amplifier and a digitizer with an built-in down-converter, before being recorded. 
To enable real-time monitoring of the signal chain, an auxiliary spectrum analyzer (SA) was incorporated. 
A vector network analyzer (VNA) was also employed for system calibration and diagnostics through multiple signal paths. 
The overall experimental setup is shown in Fig.~\ref{fig:system_overview}.

The resonant cavity used in this experiment consisted of eight identical cells separated by metal partitions within a cylindrical structure of 128\,mm inner diameter, optimized to fully utilize the available volume inside the magnet bore. 
This 8-cell configuration supports the TM$_{010}$ resonant mode at approximately 6.0\,GHz, extending the accessible frequency range to over three times that of a conventional single-cell cavity. 
An additional key design feature is the strong inter-cell coupling via a central gap, which enables coherent signal extraction through a single antenna located at the cavity center.
The cavity has an inner height of 270\,mm, yielding a total detection volume of 3.1\,L, over which the average magnetic field is estimated to be 6.97\,T.
The cavity was fabricated from oxygen-free high thermal conductivity copper, and its unloaded quality factor was measured to be approximately 45,000 at cryogenic temperatures.

\begin{figure}
    \centering
    \includegraphics[width=\linewidth]{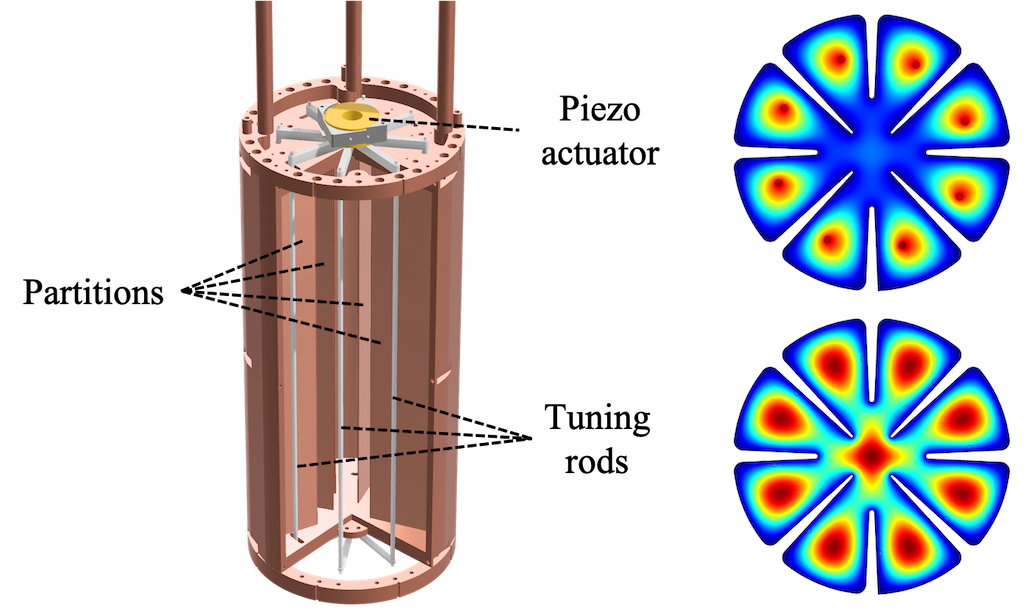}
    \caption{Left: Cutaway view of the 8-cell cavity design. Right: Simulated TM$_{010}$ electric field distributions with the tuning rod positioned near the center (top) and near the edge (bottom) of each cell, corresponding to low and high frequencies, respectively.}
    \label{fig:cavity}
\end{figure}

Frequency tuning was achieved using 3-mm-thick alumina (Al$_2$O$_3$) rods, one placed in each cell.
These rods were configured to rotate azimuthally and synchronously about the cavity center, driven by a single piezoelectric actuator mounted at the top, as illustrated in Fig.~\ref{fig:cavity}.
Finite element method simulations~\cite{COMSOL} showed that the form factor increases from 0.5 to 0.7 with frequency.
In practice, fabrication tolerances and mechanical misalignments can induce asymmetries in the electromagnetic field distribution within the cavity, leading to deviations from the ideal form factor.
To quantify this effect, we measured the electric field strength in each cell and estimated a form factor degradation of up to 5\%, depending on the position of the tuning rods.
This degradation was mitigated to below 1\% by optimizing the tuning rod diameters, as described in Ref.~\cite{JEONG2023168327}.

For quantum-noise-limited amplification, we employed a flux-driven JPA, whose fundamental design and operational principles are detailed in Refs.~\cite{APL93_042510_2008, SST34_085013_2021}.
The device consists of a $\lambda/4$ coplanar waveguide resonator terminated by a superconducting quantum interference device (SQUID), which serves as a nonlinear inductor.
To operate the JPA as a linear amplifier in the three-wave mixing regime, an external magnetic flux is applied to modulate the SQUID in conjunction with a microwave pump tone at frequency $f_p$ and power $P_p$.
Upon application of an input signal at frequency $f_s$ (the signal mode), the JPA generates an intermodulation tone at $f_i = f_p - f_s$, referred to as the idler mode. 
This nonlinear mixing process enables phase-preserving amplification of the input signal.
The resonance frequency of the JPA can be tuned by varying the magnetic flux applied to the SQUID, delivered via a superconducting coil located beneath the device.
To protect the JPA from residual magnetic fields of up to 100\,mT originating from the main magnet, we employed a three-layer magnetic shield composed of aluminum, $\mu$-metal, and niobium.
The JPA was thermally anchored to the base-temperature stage and integrated into the signal path via a circulator, as shown in Fig.~\ref{fig:system_overview}.

For optimal operation targeting a gain of 20\,dB, chosen to maximize signal amplification  near the quantum noise limit while maintaining linearity and low distortion, we constructed a look-up table (LUT) of optimal pump parameters following the measurement protocol described in Ref.~\cite{KUTLULT29}. 
At each cavity tuning step, we coarsely tuned the JPA according to the LUT entry to the cavity frequency. 
This was followed by a fine-tuning procedure, adjusting $P_p$ in increments of 0.01\,dB until the gain settled within an acceptance threshold of 0.5\,dB, all completed within 3 seconds.
The JPA gain was measured {\it in situ} using the VNA by comparing the transmission spectra obtained with the JPA turned on and off.

The system noise is commonly characterized by the equivalent noise temperature $T_{\rm sys}$, which can be decomposed as $T_{\rm sys} = T_{\rm cav} + T_{\rm rcv}$, where $T_{\rm thr}$ and $T_{\rm add}$ represent the thermal noise from the cavity and the added noise from the receiver chain, respectively.
The noise temperature estimation was based on a comparative analysis of the power spectral density (PSD) measured with the JPA turned on and off.
Following the Johnson–Nyquist formula~\cite{PhysRev.32.97.JOHNSON, PhysRev.32.110.NYQUIST}, the PSDs are given by $P_\mathrm{on} = k_B G_\mathrm{on} T_{\rm sys}\Delta\nu = k_B G_\mathrm{on} (T_\mathrm{cav} + T_{\rm on})\Delta\nu$ and $P_\mathrm{off} = k_B G_\mathrm{off} (T_\mathrm{cav} + T_{\rm off})\Delta\nu$, where $G$ denotes the system gain and $\Delta\nu$ is the measurement bandwidth.
$T_{\rm off}$ was measured separately over a wide range at the initial stage of the experiment using the Y-factor method with the designated noise source, as illustrated in Fig.~\ref{fig:system_overview}.
Using $G^{\rm JPA} = G_\mathrm{on}/G_\mathrm{off}$, we obtain the system noise temperature as
\begin{equation}
	\label{eq:noise_temperature_estimation}
	T_{\rm sys} = \frac{1}{G^{\rm JPA}}\frac{P_\mathrm{on}}{P_\mathrm{off}}(T_\mathrm{cav} + T_{\rm off}).
\end{equation}
This approach requires the stability of both $G^{\rm JPA}$ and $T_{\rm off}$ throughout the experiment. 
To ensure this, we continuously monitored these parameters over 24 hours, well exceeding the data acquisition (DAQ) time at a given search frequency, and found the statistical uncertainties in $G^{\rm JPA}$ and $T_{\rm off}$ to be below 1\% and approximately 2\%, respectively.
Aggregating data from all tuning steps, the system noise temperature was found to range between 380\,mK and 500\,mK, with step-to-step variations within 10\,mK.

DAQ was carried out from December 28, 2021, to July 19, 2022, with a total downtime of approximately two months due to system warm-ups for maintenance.
The frequency range from 5.83 to 5.94\,GHz was scanned with an average tuning step of 17.4\,kHz. 
At each step, initial calibration was performed, beginning with the adjustment of the antenna coupling to a typical value of 2, followed by the characterization of cavity properties such as $\nu_c$ and $Q_L$. 
The JPA was then tuned to the cavity frequency following the aforementioned procedure. 
Subsequently, the system noise temperature was estimated using Eq.~\ref{eq:noise_temperature_estimation}. 
Upon completion of these calibrations, DAQ was conducted.
The signal from the cavity was first amplified by the JPA and a pair of HEMTs within the cryogenic system. 
The signal was further amplified at room temperature and down-converted to an intermediate frequency (IF) of 10.7\,MHz. 
The IF time-series signal was digitized and processed via a Discrete Fourier Transform to produce power spectra with a resolution bandwidth of 62.5\,Hz over a 1-MHz span.
For each frequency step, a total of 85,200 and 88,000 spectra were collected for the lower and upper halves of the scan range, respectively, which were split due to mixing with a TE mode, and then averaged into five grouped spectra.
The total DAQ efficiency, accounting for time spent on calibrations and characterization, was estimated to be 92\%.

Data analysis proceeded in four stages: preprocessing, sideband summing, spectrum combination, and hypothesis testing.
In the preprocessing stage, each averaged raw power spectrum was cleaned by excluding a 25\,kHz bandwidth centered at $f_p/2$, four times the expected axion bandwidth ($Q_a \sim 10^6$), covering 99.9\% of the potential signal power, to eliminate the interference from the pump tone and simplify the analysis.
The baseline of the remaining spectrum was estimated using a Savitzky–Golay filter~\cite{savitzky1964smoothing} with a polynomial degree of 4 and a window size of 1921 (corresponding to 120\,kHz), sufficiently larger than the axion signal bandwidth but smaller than the cavity half-bandwidth.
Each spectrum was then normalized by this baseline and shifted by unity to yield the relative power excess in each frequency bin, denoted by $\delta$. 
The distribution of $\delta$ was consistent with the standard normal distribution, validating the baseline removal procedure and confirming statistical independence among frequency bins.

\begin{figure}
\centering
    \includegraphics[width=0.9\linewidth]{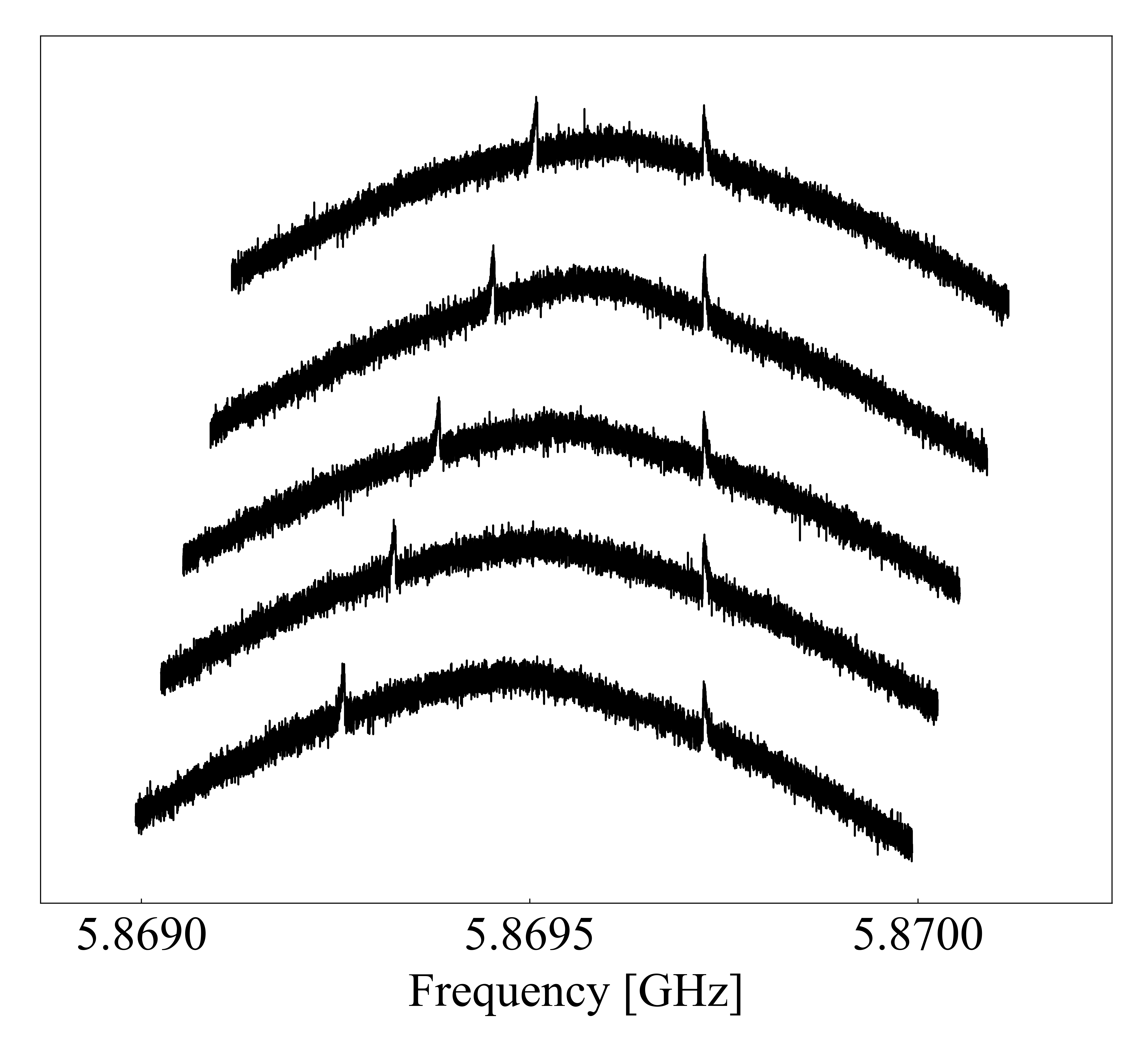}
    \caption{Power spectra with a synthetic axion signal injected at 5.86972\,GHz over multiple tuning steps. 
    The idle signals appear symmetrically with respect to the JPA center at $f_p/2$.
    }
    \label{fig:signal_overlap}
\end{figure}

A distinctive feature of our analysis is the incorporation of idler modes generated by the JPA as supplementary spectral information to enhance sensitivity. 
For each preprocessed spectrum, the analysis band was centered at $f_p/2$, approximately aligned with the cavity resonance frequency. 
A potential axion signal and its JPA-generated idler counterpart appear at frequency bins symmetrically distributed around $f_p/2$, as illustrated in Fig.~\ref{fig:signal_overlap}.
We implemented a sideband-summing technique that optimally combines these power excesses with the JPA-induced correlation accounted for.
The sideband-summed excess is defined as,
\begin{equation}\label{eq:sbexcess}
	\delta_{ss,k} = \frac{w_k\delta_{k} + w_{-k}\delta_{-k}}{\sqrt{w_k^2 + w_{-k}^2 + 2\rho w_{k}w_{-k}}},
\end{equation}
where $\delta_k$ and $\delta_{-k}$ are the power excesses in the signal and its idler bins, respectively, with optimized weighting factors $w_{k}$ and $w_{-k}$ that account for differences in the expected signal-to-noise ratio (SNR) between the two. 
$\rho$ quantifies the measured correlation between them. 
The covariance of the correlated power excesses was estimated from all individual spectra and used to determine the correlation coefficient.
A detailed discussion of the sideband-summing technique can be found in Appendix C of~\cite{PhysRevX.14.031023}.
Applying this combination across all frequency steps yielded a consistent SNR improvement of about 4\% compared to conventional analysis methods.

To proceed with data combination, we first align the sideband-summed excess in the radio-frequency domain, then combine the data in each frequency bin using weighting factors proportional to the expected SNR of hypothetical KSVZ axion signals.
From the combined spectrum, we performed a weighted sum over 25\,kHz windows at each frequency bin, using weights determined by the expected SNR based on the laboratory-frame dark matter signal lineshape~\cite{PhysRevD.42.3572.TURNER}, thereby constructing the so-called grandspectrum.
We observed several strong negative excesses deviating from the expected Gaussian distribution, particularly in the 5.87--5.88\,GHz range, similar observations reported in~\cite{PhysRevX.14.031023, PhysRevLett.134.151006.HAYSTAC}.
Although the origin of these features remains unknown, a follow-up supplementary scan with twice the statistics showed no such excesses.
Accordingly, we excluded the anomalous bins from the original preprocessed spectra and substituted them with data from the supplementary scan.

The resulting grandspectrum exhibited a standard deviation slightly below unity, a typical consequence of bin-to-bin correlations induced by the SG filter. 
The observed width of 0.85 agreed with the analytical prediction described in~\cite{Yi2023-rx}. 
For convenience of statistical interpretation, the grandspectrum was normalized to a standard normal distribution using its observed mean and standard deviation.
To estimate signal degradation caused by the SG filter, software-synthesized axion signals of various masses were injected into the data. 
The signal-injected data were then analyzed using the same procedure described above. Comparing the injected and recovered SNRs revealed an average SNR efficiency of 84\%, with only minor variations across the frequency range.

\begin{figure*}
\centering
    \includegraphics[width=.95\textwidth]{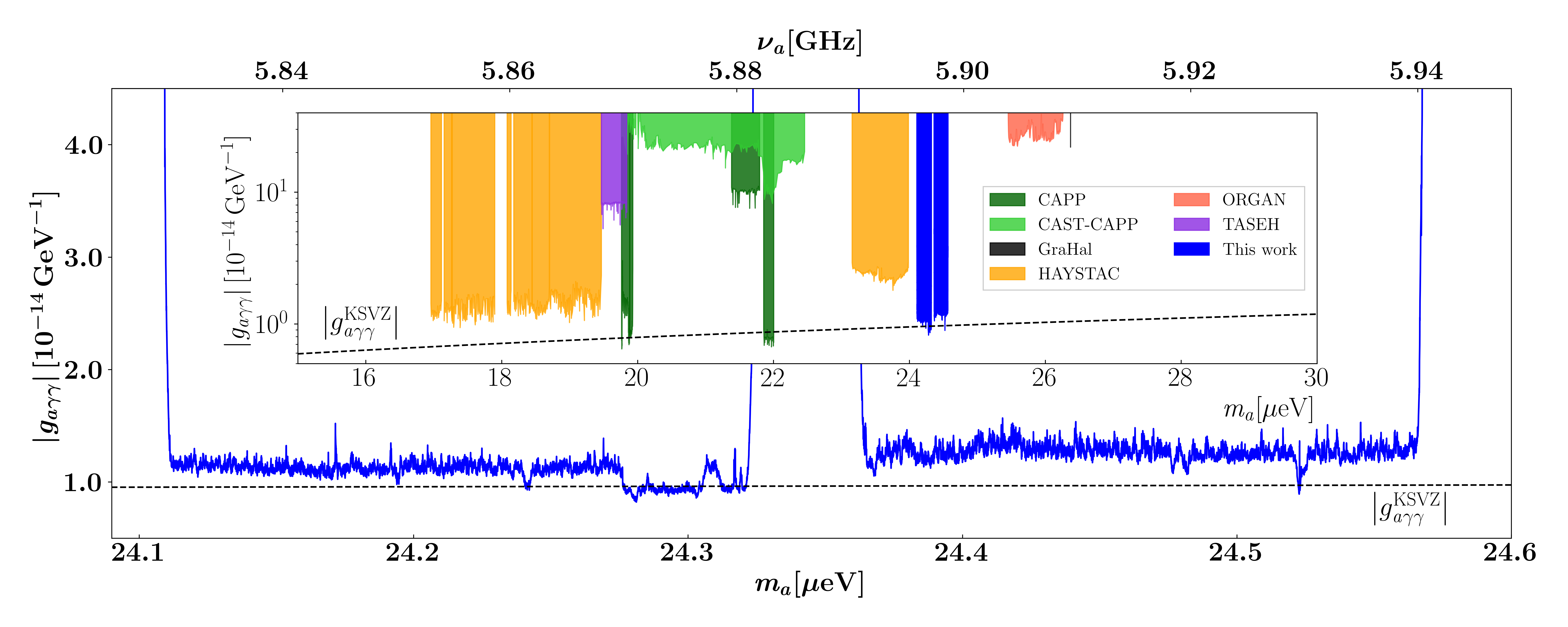}
	\caption{Exclusion limit on the axion-photon coupling, $g_{a\gamma\gamma}$, as a function of axion mass, $m_a$, set by this work at the 90\% CL. 
    The black dashed line represents the KSVZ benchmark model. 
    The gap around 24.35\,$\mu$eV is attributed to mixing with a TE mode, as confirmed by simulations.
    The inset shows the broader exclusion landscape~\cite{CAPP1,CAPP2,CAPP3,CAPP4,CASTCAPP,HAYSTAC1,HAYSTAC2,HAYSTAC3,HAYSTAC4,TASEH} in an axion mass range from 15 to 30\,$\mu$eV.
    }
    \label{fig:exclusion}
\end{figure*}

For hypothesis testing, we adopted a null hypothesis that assumes the presence of a KSVZ axion signal with ${\rm SNR} = 5$. 
Defining a detection threshold at $3.47$ for 90\% confidence level (CL), 
we identified 85 candidate bins exceeding the threshold, typically appearing in small clusters.
To evaluate the nature of these excesses, we performed follow-up scans at the corresponding frequencies with higher integration time. 
None of the candidates exhibited excess power consistent with a persistent signal, indicating they were statistical fluctuations.
In the absence of a confirmed signal, we set a 90\% CL upper limit on the axion-photon coupling, reaching $g_{a\gamma\gamma} < 1.2 \times 10^{-14}$\,GeV$^{-1}$ over the axion frequency (mass) range $5.82 < f_a < 5.93$\,GHz ($24.11 < m_a < 24.57\,\mu$eV).
This corresponds to the most stringent limit to date in this mass range, achieving sensitivity near the KSVZ benchmark prediction.
The resulting exclusion limit is presented in Fig.~\ref{fig:exclusion}. 

In conclusion, we report a search for axion dark matter using a novel 8-cell microwave cavity designed to increase frequency coverage while preserving detection volume, combined with a state-of-the-art flux-driven JPA operating near the quantum noise limit.
A sideband-summing technique was employed to coherently combine symmetric spectral features from the JPA’s signal and idler modes, resulting in a 4\% improvement in SNR.
Over a 100-MHz frequency range near 5.9\,GHz (24.4\,$\mu$eV), we exclude axion-photon couplings $g_{a\gamma\gamma} \gtrsim 1.2\times10^{-14}\,{\rm GeV^{-1}}$ at 90\% confidence level, approaching the sensitivity of the KSVZ benchmark model.
Future efforts will extend the search to adjacent axion frequencies with sustained sensitivity, utilizing alternative cavity and amplifier configurations.

This work was supported by the Institute for Basic Science (IBS-R017-D1 and IBS-R040-C1) and JSPS KAKENHI (Grant No.~JP22H04937).
J. Jeong was partially supported by the Knut and Alice Wallenberg Foundation.
A. F. Loo was supported by a JSPS Postdoctoral Fellowship.
J. E. Kim was partially supported by the Korea Academy of Science.

\bibliography{capp8tb}
\end{document}